\documentclass[preprint]{aastex}
\begin{document}
\title{CREATION OF SPIRAL GALAXIES II}
\author{Masataka Mizushima}
\affil{Department of Physics, University of Colorado,\\Boulder, Colorado 80309}
\begin{abstract}
        The spiral galaxies, including our galaxy, are created by the
	gravito-radiative forces generated by colliding black holes
        at the center of quasars. The gravito-radiative force is
	predicted by Einstein's general theory of relativity. A
	quasar is assumed to have a circular disk of highly
        compressed neutrons (ylem) orbiting around black holes.
	The collision of two black holes at the center generates the
	gravito-radiative force, exerted on the ylem disk, producing
	a pair of bars of stars with $180^{\circ}$ rotational
        symmetry. This pair of bars develop into a pair of spiral arms,
        keeping the $180^{\circ}$ rotational symmetry.
	Therefore, the number of spiral arms must be even. Our Milky
        Way galaxy has two pairs of arms, and has the 180$^{\circ}$
        rotational symmetry, indicating that we have had two
	galactic nuclear explosions. The theory proposed by Gamow
	and others on the making of chemical elements fits into this
	theory.  Thus, the age of the Milky Way galaxy must be equal
        to or greater than the age of the earth, 4.5$\times10^{9}$ yr.
	The spirality of the Milky Way galaxy is examined under this
	assumption, and it is found that our galaxy was once about
	10 times larger than it is now, and has been shrinking
	during the last half of its life.
\end{abstract}
\maketitle{}

\pagebreak
\section{QUASARS}
\indent Hubble found that the estimated distance of a galaxy from us
is proportional to the recession velocity of that galaxy from us.
The proportionality constant, called Hubble's constant, $H$,
is about 77 km/s per Mpc, or 3 km/s per 10$^{21}$m.
Thus the universe is expanding.
The distance of galaxies from us, however,
has an upper limit of about $d_{max}$ = 3000 Mpc, 
and most of the quasars are found beyond that distance with
the same Hubble's constant \citep{1} as seen in Fig.1. 
\begin{figure}[ht]
\begin{center}
\includegraphics[width=9cm]{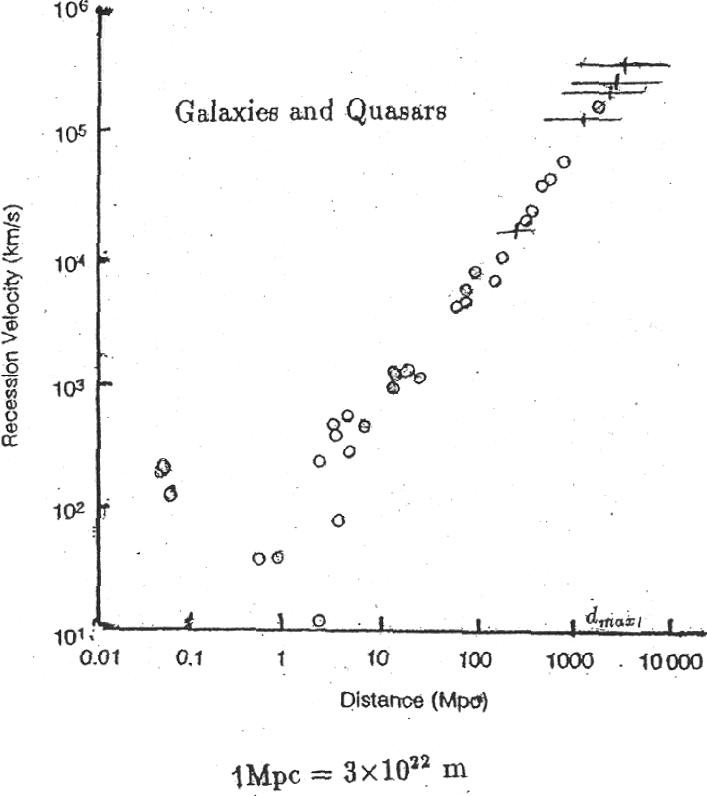}
\caption{Estimated distances of galaxies
and quasars \citep{1}. Galaxies are shown by $\circ$, and quasars are shown
by +. The distance is proportional to the recession
velocity, showing Hubble's law. Almost all
galaxies exist only within the distance,
$d_{max}$, such that the corresponding time,
$d_{max}/c$,  is about $10^{10}$ years.}
\end{center}
\end{figure}

\indent Each quasar is regarded as an active galactic nucleus without
spiral arms, but we do not know its internal structure \citep{2}. We assume that
esch quasar has some black holes and a highly condensed neutron (ylem) disk is
orbiting around them. This disk developed into a galaxy at
$d_{max}/c = 10^{17}$ s = $10^{10}$ years ago. That is, almost all galaxies
were created at about the same time, here.\\
\indent The age of the earth is geologically known to be $4.5\times10^{9}$ years,
and the earth is probably created shortly after the Milky Way galaxy was
created. Thus, we can assume that our galaxy was created from a quasar, among
all other galaxies, $10^{10}$ years ago. When and how the quasars were created
is another problem \citep{4}.\\
\section{GRAVITO-RADIATIVE FORCES}
\indent Einstein assumed that
\begin{equation}
	\renewcommand{\theequation}{2-1}
	\left(\frac{dx}{dt}\right)^{2} =
	c^{2} - \left(\frac{ds}{dt}\right)^{2} \leq c^{2},
\end{equation}
for a particle moving in the $x$-direction, and formulated the
special theory of relaitivity. He generalized this assumption into a
metric
\begin{equation}
	\renewcommand{\theequation}{2-2}
	ds^{2} = g_{00}(ct)^{2} - g_{xx}(dx)^{2}
	- g_{yy}(dy)^{2} - g_{zz}(dz)^{2},
\end{equation}
and proposed that gravity could be described in terms of the space-time
dependence of the metric coefficients, $g_{ij}$. Therefore, in this theory,  
the general relativity theory, gravity also propagates with a finite
speed equal to or less than the speed of light.\\ 
\indent In Einstein's theory, equations of motion under gravitational
fields can be obtained from the generalized Hamilton's variational principle,
\begin{equation}
\renewcommand{\theequation}{2-3}
\delta\int ds = 0,
\end{equation}
as 
\begin{equation}
	\renewcommand{\theequation}{2-4}
	\frac{d^{2}x^{i}}{ds^{2}} =
	-\Gamma^{i}_{k\ell}\frac{dx^{k}}{ds}\frac{dx^\ell}{ds}
	\equiv F_{E}^{i}/(mc^{2}),
\end{equation}
where the $\Gamma^{i}_{k\ell}$ are called Christoffel symbols.
Because the space component of the left-hand side of
eq.(2-4) reduces to the acceleration in $c^{2}$,
we call the space component of the right-hand side of 
eq. (2-4) the Einstein force ${\bf F}_{E}$ in $mc^{2}$. The wave equations of
Einstein, when linearized, that is, when the higher order terms of the 
deviation of the metric components from 1 are neglected show \citep{5} that these 
metric components can be expressed in terms of gravitational scalar, vector, 
and tensor potentials, 
\begin{equation}
\renewcommand{\theequation}{2-5}
\phi_{00}({\bf r},t) = \frac{4GM}{c^{2}r},~~~
\mbox{\boldmath$\phi$}({\bf r},t) = - \frac{4GM{\bf v}}{c^{3}r},~~~{\rm and}~~~
\phi_{\alpha\beta}({\bf r},t) = \frac{4GMv_{\alpha}v_{\beta}}{c^{4}r},
\end{equation}
respectively, where $\alpha, \beta$ = 1, 2, and 3. We neglected the retardation 
effects, which are not important in our cases. In the linear approximation, 
equation (2-4) can be written as \citep{7} 
\begin{equation}
\renewcommand{\theequation}{2-6}
\frac {d^{2}{\bf r}}{dt^{2}}  
= -\frac{c^{2}}{4}{\bf\nabla}\phi_{00} 
- c\frac{\partial{\mbox{\boldmath$\phi$}}}{\partial t}
+ c{\bf V}\times({\bf\nabla}\times{\mbox{\boldmath$\phi$}})
+ \frac{\partial [{\bf v}({\bf V}\cdot{\mbox{\boldmath$\phi$}})]}{c^{2}\partial t}
\equiv \frac{{\bf F}_{E}}{m}
\end{equation}

where ${\bf v}$ and ${\bf V}$ are the velocities of a source particle
(of mass $M$) and a test particle of mass $m$, respectively.
In the above expression,
we do not differentiate the velocity of the test particle, ${\bf V}$,
by time $t$.\\
\indent In eq. (2-6), we see that the Newtonian term appears as the first
approximation, as expected, but there are other correction terms due to
Einstein's principle, which says that gravity propagates with the speed
of $c$. We notice that there are two terms
in eq. (2-6) that are proportional to $1/r$ instead of $1/r^{2}$, as the
Newtonian term is. The Newtonian term gives the orbit of a test particle
to the first approximation, but the two terms proportional to $1/r$
transmit energy and momentum through space, as the Poynting theorem shows
in the theory of electricity and magnetism. These two terms are
\begin{equation}
\renewcommand{\theequation}{2-7a}
{\bf F}_{rad1}/m = \frac{4GM{\dot{\bf v}}}{c^{2}r},
\end{equation}
and
\begin{equation}
	\renewcommand{\theequation}{2-7b}
	{\bf F}_{rad2}/m =
	-\frac{4GM\left[{\dot{\bf v}}({\bf V}\cdot{\bf v}\right)
	+{\bf v}({\bf V}\cdot{\dot{\bf v}}]} {c^{4}r}.
\end{equation}
These two terms are called the
gravito-radiative forces of order 1 and 2, respectively.
The ${\bf F}_{rad2}$ actually comes from the gravitational
tensor potential, but in eq. (2-6) it is expressed in terms
of the gravitational vector potential.\\
\indent The above forces appear in the geodesic deviation, but \citep{5}
missed ${\bf F}_{rad2}$ by simply assuming that ${\bf V} = 0$ for the
test particle.\\
\indent A collision between two massive black holes at the center
of a quasar generates a gravito-radiative force ${\bf F}_{rad2}$.

\section{CREATION OF ENERGY AND ANGULAR MOMENTUM}
When the radiative forces are produced by collisions of two or more massive
particles (black holes) located near the origin, Newton's third law, 
$\Sigma_{i} M_{i}{\dot{\bf v}}_{i} = 0$, is applied to null the total contribution
of $\Sigma_{i}{\bf F}_{{rad1} i}/m = 0$.\\
\indent The total energy of a test particle of mass $m$ and speed $V$ under the gravitational 
field due to mass $M$ at the origin is
\begin{equation}
\renewcommand{\theequation}{3-1}
{\cal E} =  mc^{2} + \frac{1}{2}mV^{2} - \frac{GMm}{r},
\end{equation}
to the first approximation in $(V/c)^{2}$ of static general relativity, and 
it agrees with the Newtonian theory. In the Newtonian theory, we know that
$d{\cal E}/dt = 0$,
but in the general relativity theory, assuming that $dm/dt = 0$, 
we see that
\begin{displaymath}
\frac{d{\cal E}}{dt} = m\frac{{\bf V}\cdot d{\bf V}}{dt} 
- GMm\frac{d(1/r)}{dt}
\end{displaymath}
\begin{equation}
\renewcommand{\theequation}{3-2}
= {\bf V}\cdot{\bf F}_{E} - GMm\frac{d(1/r)}{dt}
= -\frac{8GMm({\bf V}\cdot{\bf v})({\bf V}\cdot{\dot{\bf v}})}{c^{4}r},
\end{equation}
by taking ${\bf F}_{E}$ for $md{\bf V}/dt$ from eq. (2-6). Thus, an energy is 
created by ${\bf F}_{rad2}$ at the collision of black holes and transmitted to
the ylem orbiting around the center.\\
\indent \citep{6} and \citep{8}
derived a formula which gives the rate of
energy loss of mass $M$ by emitting the gravitational wave into space, and
their result was confirmed by the observatin on the Binary Pular PSR-1913+16
\citep{9}. Our equation (3-2) gives the corresponding rate of the energy gain of
mass $m$ by receiving from the same gravitational wave.\\
\indent We assume that the ylem form a circular disk orbiting around the
center, and that the collision of two black holes takes place head-on in the 
ylem's plane. Equation (3-2) shows that only those ylems that are orbiting
on two sides of a circle at which ${\bf V}$ is parallel to ${\bf v}$ or
${\dot{\bf v}}$ would gain energy from the gravito-radiative force. The ylem 
orbiting on the part of the circle between these two parts are not touched
\citep{11}.\\
\indent The collision must be extremely relativistic. The velocity $v$ must
be close to $c/2$, and $\Delta t {\dot v}$ must also be close to c/2, where 
$\Delta t$ is the duration time of the collision. Thus, the magnitude of the energy
that part of the ylem disk (with mass $m$) gained must be about
\begin{equation}
\renewcommand{\theequation}{3-3}
\Delta{\cal E} \simeq \frac{2GMm({\bf V}\cdot{\bf n})^{2}}{c^{2}r},
\end{equation}
where ${\bf n}$ is a unit vector in the direction of the (head-on) collision
of black holes.\\ 
\indent If $r_{\circ}$ is the black hole radius, the speed of the 
ylem, $V$, orbiting at $4r_{\circ}$ from the center of the quasar, must be 
close to $c/2$. The extra kinetic energy for each of these two closest parts of 
the ylem disk that is orbiting in the direction of ${\bf n}$, or -${\bf n}$ 
gains, is about $\Delta{\cal E} \simeq GMm/r$, that is, 
just enough to send it out to the edge of the galaxy. If the ylem disk were orbiting
around the black holes following Newtonian dynamics, the orbiting speed $V$
of a part of the disk at a distance $r$ from the center must be proportional
to $1/\sqrt{r}$. Thus, those ylems near the rim of the original disk would
gain energy much less than the above-stated limitimg amount. The two parts of 
the disk would develop into two bars continuous down to the radius of the 
original quasar. Because they gain angular momentum at the time of explosion, 
these two bars develop into two spiral arms.\\ 
\indent If the original ylem disk was  
circular, the resulting two arms would be symmetric with a 180$^{\circ}$
rotation around the center. Most of the spiral galaxies observed have
that feature, and the rest show only their flat side view. There may have been
more than one black hole collisions at the center of
some galaxies, in which case the galaxy would have an even number of spiral 
arms with the 180$^{\circ}$ rotational symmetry. We see that galaxies
NGC2997, NGC3310, M33, M51, and M101 have two pair of arms with the
180$^{\circ}$ rotational
symmery. Galaxy NGC 7137 is reported to have three arms. It is possible
that a black hole collision had a strange orientation to violate the
symmetry in that galaxy.\\
\indent Our Milky Way galaxy has two pair of spiral arms, in which the $180^{\circ}$ rotational 
symmetry is maintained \citep{13}. They are named (Sagittarius-Cygnus) and 
(Scutum-Perseus). We have had two black hole collisions in this galaxy.\\
\indent Every spiral galaxy, except for NGC 7137, belongs to the
crystallographic point group $C_{2h}$.\\
\indent The neutrons which used to form the ylem, now became free out of the 
strong gravitational compression, and started the nuclear reaction, 
$n \rightarrow p + e + \nu$, producing electro-magnetic fields and chemical
elements. However, the galactic explosion ended too soon, and the synthesis
of chemical elements ended unfinished, as Gamow found \citep{14}. The electro-magnetic
radiation of 3K remained as a cosmic background \citep{15}. Earth was born at the 
same time, or shortly after the Milky Way galaxy was born. Therefore,
the age of the Milky Way galaxy, or its Sagittarius arm, must have an age almost
equal to the age of the earth, $4.5\times10^{9}$ years.
\section{SPIRALITY OF MILKY WAY GALAXY}
Our galaxy, the Milky Way galaxy, has two pairs of spiral arms, and has a
180$^{\circ}$ rotational symmetry, confirming our theory of its 
creation. Following one of the initial explosions, a pair of arms is
extended radially out of the core, and each star in each arm starts orbiting
around the center of the galaxy due to the torque given at the initial 
explosion, as explained in section 3.\\
\indent The motion of each star, after the initial explosion, must follow
the Newtonian theory. But we do not know the distribution of masses, including
the dark matters, during that period.
We may, however, assume that the effective gravitational
forces are mostly in the radial direction, and the tangential motion can
be discussed separetely.\\ 
\indent At a given distance, $r_{1}$, from the center, the stars in a spiral
arm at this distance travel with a speed $V_{1}(t)$ tangentially at
time $t$. Each star at this distance covers an angle $\theta_{1}$ given by 
$\int V_{1}(t)dt/r_{1}$, the integral of $V_{1}(t)$ over the entire lifetime,
if it has been moving tangentially, starting from
a straight line given by the initial explosion. Thus, taking another radial
distance $r(2)$ from the center of the galaxy
\begin{displaymath}
\int(V_{1}(t) - V_{2}(t))dt = (<V_{1}> - <V_{2}>)\Delta t_{a} 
\end{displaymath}
\begin{equation}
	\renewcommand{\theequation}{4-1}
	= \theta_{1}r_{1} - \theta_{2}r_{2}
	= \frac{1}{2}(\theta_{1} - \theta_{2})(r_{1} + r_{2})
	+ \frac{1}{2}(\theta_{1} + \theta_{2})(r_{1} - r_{2}),
\end{equation}
where $<V>$ is the average of $V$ over the age $\Delta t_{a}$ of the galaxy.
In the present Milky Way galaxy, we know that all stars are orbiting around 
the center with a tangential speed, $V_{tan}$, of about 220 km/s, and this 
speed is almost independent of the radial distance from the galactic center,
within the accuracy of about 10 km/s. 
Let us take the Sagittarius arm of the Milky Way galaxy, as an example. 
If we take positions of stars $\pi/2$ from that of the sun on each side as
$r_{1}$ and $r_{2}$, we see \citep{13} that $r_{1}$ = 10 kpc and $r_{2}$ = 5 kpc.
Therefore, eq. (4-1) gives
\begin{equation}
\renewcommand{\theequation}{4-2}
	<V_{1}> - <V_{2}> =
        -4.5\times10^{3} + 0.5\times10^{3}(\theta_{1} + \theta_{2}).
\end{equation}
in m/s, if we take $\Delta t_{a} = 4.5\times10^{9}$
yr = $1.5\times10^{17}$s, the age of Earth.
This relation, eq. (4-2), is consistent with the present situation, 
where $V_{1} \simeq V_{2}$, if $(\theta_{1} + \theta_{2})$ = 10.
On the other hand
\begin{displaymath}
<V_{1}> + <V_{2}> = \frac{1}{2\Delta t_{a}}(\theta_{1} - \theta_{2})(r_{1} - r_{2})
+ \frac{1}{2\Delta t_{a}}(\theta_{1} + \theta_{2})(r_{1} + r_{2})
\end{displaymath}
\begin{equation}
\renewcommand{\theequation}{4-3}
= 1.5\times10^{3}(\theta_{1} + \theta_{2}) + 1.5\times10^{4},
\end{equation}
using $\Delta t_{a} = 1.5\times10^{17}$ s. If we take $(\theta_{1} 
+ \theta_{2})$ = 10 and we obtain $<V_{1}> = <V_{2}>$ = 15 km/s.\\
\indent The assumption that $V_{1} = V_{2}$ during the entire life of the
Sagittarius arm may not be correct. In fact the inaccuracy of this assumption at present
is about 10 km/s. Thus, $\theta_{1} + \theta_{2}$ in eq. (4-2) can be as large
as 30. If we take this value in eq. (4-3) we obtain $<V_{1}>$ as large as 35
km/s, instead of 15 km/s, but 35 km/s may be the largest value possible.
At present, we know that the tangential speed of stars is 
about 220 km/s, and almost independent of the distance from the galactic 
center. Therefore, we conclude that the average tangential speed in the
past was about 0.15 of the present value.\\
\indent If the gravitational forces were radial, the orbital angular momentum
must have been conserved after the
galactic nuclear explosion. If the tangential speed $V$ has been
independent of the distance from the center during the past, then
$V\ell_{\circ}$ is conserved, where $\ell_{\circ}$ is the length of
the spiral arm. Therefore, 
\begin{equation}
\renewcommand{\theequation}{4-4}
\ell_{\circ,p}/<\ell_{\circ}> = <V>/V_{p} = 0.15,
\end{equation}
where $V_{p}$ = 220 km/s and $\ell_{\circ,p}$ which is $\ell_{\circ}$ 
at present. Thus.
we conclude that the length of the spiral arm was about 7 times the present
length on average during the past. That is, the size of the galaxy was
at least 10 times that of the present size, at one time \citep{18}.\\
\indent This also means that the galaxy is now shrinking at the 
rate of about 10 km/s, which is below the present accuracy of observation.
The Milky Way galaxy is the oldest galaxy.\\
\acknowledgments{I thank Professor David Bartlett for reading this paper
critically and showing me new data.}

\end{document}